# NN2CAM: Automated Neural Network Mapping for Multi-Precision Edge Processing on FPGA-Based Cameras

Petar Jokic, *Member, IEEE*, Stephane Emery, *Member, IEEE*, and Luca Benini, *Fellow, IEEE*

*Abstract*—The record-breaking achievements of deep neural networks (DNNs) in image classification and detection tasks resulted in a surge of new computer vision applications during the past years. However, their computational complexity is restricting their deployment to powerful stationary or complex dedicated processing hardware, limiting their use in smart edge processing applications.

We propose an automated deployment framework for DNN acceleration at the edge on field-programmable gate array (FPGA)-based cameras. The framework automatically converts an arbitrary-sized and quantized trained network into an efficient streaming-processing IP block that is instantiated within a generic adapter block in the FPGA. In contrast to prior work, the accelerator is purely logic and thus supports end-to-end processing on FPGAs without on-chip microprocessors. Our mapping tool features automatic translation from a trained Caffe network, arbitrary layer-wise fixed-point precision for both weights and activations, an efficient XNOR implementation for fully binary layers as well as a balancing mechanism for effective allocation of computational resources in the streaming dataflow. To present the performance of the system we employ this tool to implement two CNN edge processing networks on an FPGA-based high-speed camera with various precision settings showing computational throughputs of up to 337GOPS in low-latency streaming mode (no batching), running entirely on the camera.

## I. INTRODUCTION

A wide range of deep neural networks (NN) are being used today to detect and classify objects [1] in imaging applications but also serve many other domains that go far beyond the field of computer vision. The list of different network architectures is nearly as long as the number of applications itself, each one optimized for its specific purpose, the hardware it is running on, the available power, and the computational throughput that is expected. To efficiently design, train, and deploy each one of them, a diverse set of strategies have been followed. Thus, some networks rely on accurate high-precision computations while others are optimized for fast, but lower-precision, computations [2]. For an efficient and quick deployment of a NN in an embedded system it is therefore crucial that the processing hardware and its mapping tool support the optimal parameter options.

What all, and especially the deep, NNs have in common is their computational complexity, requiring powerful computational resources that allow the processing of billions of multiply-and-accumulate (MAC) operations at low latency to meet the throughput requirements of the application. While such a workload is usually too demanding for general-purpose central processing units (CPU), the highly parallelized architectures of graphics processing units (GPU) offer sufficient computational throughput for processing small- to medium-sized networks in real-time [3]. Thus, most of today's machine learning frameworks for training and inference directly support GPUs for accelerating the processing. Cloud computing solutions are employed in some use-cases, outsourcing the large and power-hungry processing engines to servers with virtually unlimited resources. While this provides mobile systems with access to cheap and powerful computational resources, it introduces energy-intensive (raw) data communications to the cloud and adds significant latency to the processing.

For applications like video self-triggering, ensuring low processing latency is crucial as it was shown in [4] and [5]. These implementations achieve low-latency video triggering through on-board image comparison or temporal change detection but could be extended with more sophisticated frame-based triggering algorithms using on-board processing capabilities. This so-called *edge processing* enables low-latency without consuming large amounts of power on transferring data to the cloud and back, which would additionally expose (possibly sensitive) data to the network and thus introduce privacy concerns.

The general-purpose architectures of CPUs and GPUs enable the user to easily adapt to new NN architectures, which might require novel data types, connection types, and other layer dimensions. However, each specific network can be

Petar Jokic is with the Swiss Federal Institute of Technology, ETH Zurich, 8092 Zurich, Switzerland and CSEM SA, 8005 Zurich, Switzerland (email: petar.jokic@csem.ch).

Stephane Emery is with CSEM SA, 8005 Zurich, Switzerland (email: stephane.emery@csem.ch).

Luca Benini is with the Swiss Federal Institute of Technology, ETH Zurich, 8092 Zurich, Switzerland and the University of Bologna, 40126 Bologna, Italy (email: lbenini@iis.ee.ethz.ch).

implemented more efficiently using application-specific integrated circuits (ASIC), which can avoid all unnecessary overhead, implementing only those hardware resources that are required for a specific target application. This minimizes both the memory footprint and the power consumption for processing the network, making them suitable for edge processing on low-power and low-cost devices. Unfortunately, due to the fast evolution of neural networks in the past few years and the long development time for ASICs, the state-of-the-art (SoA) network types might already have significantly changed by the time the ASIC arrives on the market.

The resulting need for more flexible hardware accelerators opened the door for FPGA-based devices, providing a wide range of highly configurable hardware blocks that can quickly be reconfigured to optimize the allocation of computational resources to the new network architectures. The time-consuming programming flow of FPGAs using hardware description language (HDL) can be streamlined with high-level synthesis (HLS) tools like Xilinx Vivado HLS [6]. Furthermore, a range of domain-specific NN mapping tools have been presented to simplify the implementation of trained NNs on FPGA-based devices. They support a variety of features like parameter quantization, computation scheduling, automatic extraction of parameters, and most importantly the implementation of the accelerator on the FPGA. However, these tools often rely on vendor-specific hardware blocks and are limited to a subset of the available FPGA chips (or series), as they require hard CPU cores integrated within the FPGA fabric [2], [7]. Other tools are further restricted to specific network parameters (e.g. 3x3 kernels only [8]) or supported numerical precision (e.g. binary precision only [2] or floating-point only [7]). Furthermore, most of these tools rely on batching to achieve high utilization at the expense of latency.

This work presents NN2CAM, a flexible and automated framework for mapping NNs as highly parallelized low-latency, streaming implementations onto pure-logic FPGA platforms for edge processing. Our target platform is a FastEye [9] high-speed camera as shown in Fig. 1. The tool automatically extracts the network architecture and its parameters from a trained network, supporting layer-wise configurable fixed-point precisions of weights and activations with an efficient XNOR-implementation for fully binary layers. Each network is implemented as a streaming architecture, reducing the memory needs and enabling parallel computations on all layers simultaneously. A resource-balancing technique ensures that the available processing hardware resources are effectively distributed among the layers to maximize the throughput. In section II we give a brief introduction to neural networks and processing approaches. Section III provides an overview of existing SoA frameworks for mapping networks onto FPGAs and a comparison with our approach. In section IV we present our proposed framework for mapping neural networks on FPGA-based devices and show experimental results on the high-speed camera platform in section V.

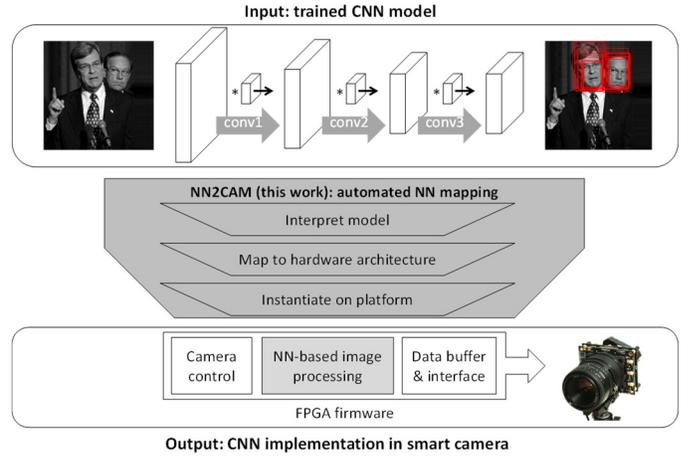

Fig. 1 Overview of the NN2CAM framework, mapping a trained CNN model onto a smart FPGA-based camera.

## II. BACKGROUND

Neural networks exist in many (application-dependent) forms, mainly differing in their dimensions, layer types (e.g. convolution layers), and the data precision. To deploy a NN onto a hardware platform, the mapping tool must understand these abstract parameters and be able to create an (efficient) mapping between the network description and the available hardware resources. In the following subsections, we summarize the basic concept behind neural networks and define the terminology for interpreting neural network models.

While biological neurons operate in an analog fashion, most artificial neural networks work with digital data, requiring a well-defined numerical representation of the underlying information. The level of quantization in these representations largely influences the accuracy of a network but also impacts the computational complexity of the required processing hardware. Thus, we further elaborate on quantized networks in the second subsection. We note that artificial NNs can also be computed in the analog domain, trying to mimic the biological world more closely, which however goes beyond the scope of this work. In the last subsection, we present two processing approaches, namely streaming and layer-wise processing, describing in what order the network is processed from its input to the resulting output layer.

### A. Neural Networks

Neural networks were inspired by the biological structure of the human brain [10], consisting of neurons that receive input activation signals $a_{i,n}$ from other neurons, weigh them with a scaling factor $w_{n,k}$, accumulate the weighted input signals, and pass the sum through an activation function $f_{act}$, resulting in an output activation signal $a_o$ as shown in equation (1). This formula can be interpreted as a dot-product between the two vectors $a$ and $w$. The most often used activation function is the rectified linear (unit) function (ReLU), shown in equation (2).

$$a_{o,n} = f_{act}\left(\sum_{k \in N_{in}} a_{i,k} \cdot w_{n,k}\right) \quad (1)$$
$$f_{act,ReLU}(x) = \max(0, x) \quad (2)$$

Neurons in NNs are usually grouped into layers that are connected to neighboring layers, forming a so-called multi-layer perceptron [10]. While the first and last layers are usually accessible by an application (inputs, results), the layers in between are called hidden layers and hold extracted features (thus, they are often called feature maps). Inter-layer connectivity differs across layer-types: in fully-connected (FC) layers each neuron of a layer receives inputs from all neuron outputs in the previous layer, while each neuron in a convolutional neural network (CNN) only receives inputs from a small connected region on the previous layer, largely reducing the number of connections. CNNs were found to achieve high accuracy for image-based object detection and recognition tasks [11], where the first input layer is usually the (2- or 3-dimensional) image. The weights are reused for each neuron, forming a 2-dimensional convolution operation between the input feature map $in$ and a kernel $k$ containing the weights as shown in Fig. 2. The visualization shows the computation of a generic CNN layer with an input feature map of size $X_{in} \cdot Y_{in} \cdot C_{in}$ being convolved with $C_{out}$ kernels of size $K_x \cdot K_y \cdot C_{in}$ and thus resulting in an output feature map of size $X_{out} \cdot Y_{out} \cdot C_{out}$. The gray window on the input activations in Fig. 2 shows the current position of the convolution operation (where the kernel is convolved with the gray-marked activations). During the convolution, this window is slid in x- and y-direction over the input activations, performing a so-called sliding window operation. Fig. 3 shows the six nested loops that explain the CNN computation of a layer in more detail.

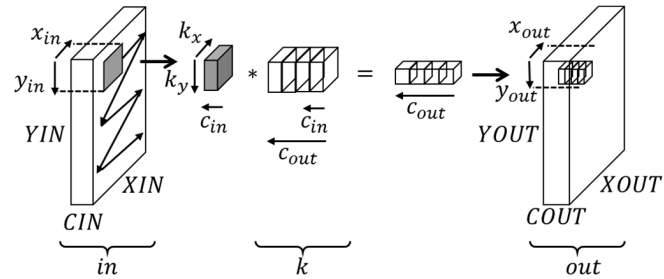

Fig. 3 Visualization of CNN processing using sliding window operation for implementing 2D convolution with a kernel.

for $y_{out}$ in 0 to $Y_{out}(i)$
    for $x_{out}$ in 0 to $X_{out}(i)$
        for $c_{out}$ in 0 to $C_{out}(i)$
            for $k_y$ in 0 to $K_y(i)$
                for $k_x$ in 0 to $K_x(i)$
                      for $c_{in}$ in 0 to $C_{in}(i)$
                          $y_{in} = y_{out} \cdot s_y(i) - P_y(i)$
                          $x_{in} = x_{out} \cdot s_x(i) - P_x(i)$
                          $out(y_{out}, x_{out}, c_{out}) \mathrel{+}= \backslash$
                              $in(y_{in} + k_y, x_{in} + k_x, c_{in}) \cdot k(k_y, k_x, c_{in}, c_{out})$

Fig. 4 Nested loops for computing a generic CNN layer (i) (the accumulation reset and activation function at the end of each $c_{out}$ loop are omitted).

CNNs are a superset of FC networks as choosing the convolutional kernel to be the size of the input feature map converts a CNN into an FC network. Thus, we will focus on CNNs in the following, noting that the same properties apply to FC layers.

### B. Quantized Neural Networks

The precision of activation and parameter values in NNs can differ across various implementations and even across different layers within a single network. While most of the powerful general-purpose processing devices operate with 32 bit floating-point numbers, allowing to accurately represent a wide range of values, power- and size-restricted platforms are often limited to smaller and fixed-point precisions ranging from 16 bit down to 1 bit values. Smaller fixed-point representations consume less memory (32x less for binary precision) for storing them and their arithmetic operations require much smaller hardware resources and related power per operation than their 32 bit floating-point counterparts [12].

Quantizing NNs to smaller precisions requires some attention as quantization errors can quickly propagate through deep networks, accumulating to significant sources of accuracy deteriorations. Research in this field has shown that aggressive quantization can be achieved with little to no impact on accuracy by taking the quantization levels into account during network training [13], [14]. Such resiliency to quantization errors has been shown for low bit precisions ranging from a few bits [15] down to binary representations [16].

### C. Streaming versus Layer-Wise Processing

Due to the layered structure of NNs, they can be computed in a layer-wise fashion, always completing one layer before starting the next one, as shown in Fig. 4 a). Once a layer is finished, the memory of its input layer is no longer needed and can be overwritten (freeing that memory space). This has the advantage that only a maximum of two subsequent activation layers must be kept in memory instead of all of them [17], reducing the memory needs. However, computation parallelism is similarly limited to a single layer, increasing the pressure on the activations memory for providing sufficient input activations to feed parallel processing units.

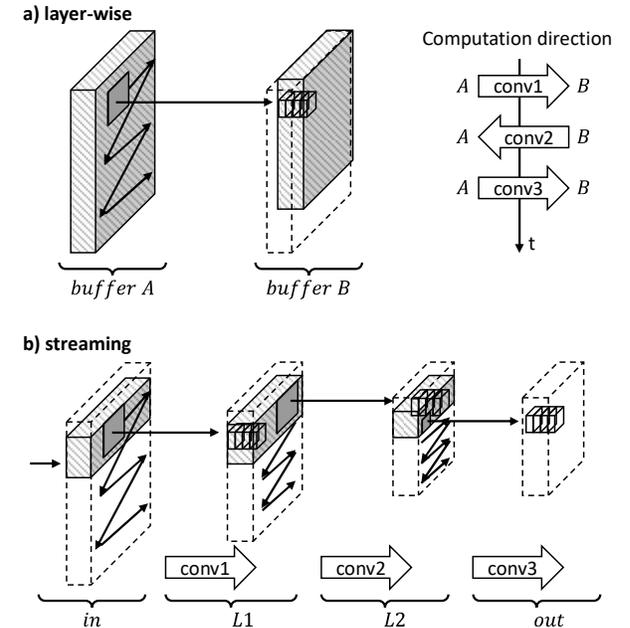

Fig. 2 Visualization of layer-wise CNN processing (a) and comparison with streaming implementation (b). The grey box marks the window position that is currently being computed (convolved with the kernel) while the striped box indicates data that is buffered in memory.

As opposed to the layer-wise computation, streaming processing (also called depth-first execution [18]) has a small portion of each layer buffered in memory. This allows activations on every layer to be processed separately as soon as sufficient data from the previous one is available. Parallelization of computations can thus be extended to many layers and the required activation memory can largely be reduced as only a small portion of each layer must be stored at any point in time. This principle is depicted in Fig. 4 b), showing the entire layer (activations) with dashed lines and the part that is stored in memory with solid lines. Each layer continuously buffers incoming activations computed by its preceding layer and processes them as soon as a complete kernel-sized window is available. Data that is no longer needed in other sliding window positions can be overwritten, making the buffer a first-in-first-out (FIFO) block. The results from each convolution with a kernel get streamed out to the next layer, which operates in the same FIFO manner.

## III. RELATED WORK

Several existing frameworks can be used to map certain types of neural networks onto FPGA platforms, each one specialized for a target field of applications and thus with a different set of supported features. Mittal [19], Guo et al. [20] and Kim [21] provide extensive surveys of previously proposed optimization strategies and implementations for FPGA-based NN accelerators. A list of SoA frameworks, along with the well-known Caffeine, DLAU, and Caffeinated FPGAs frameworks as reference, can be found in Table I.

After the trend of DNNs started in the past decade, the Caffeine framework [22] was presented to accelerate the inference of networks on a PC by offloading some computations to an FPGA, connected through a PCI interface. The software library compiles layer settings and parameters that are then synthesized using the Xilinx SDAccel tool and loaded onto the FPGA, which computes some parts of the network layer-by-layer. A similar implementation, but with a pre-synthesized accelerator that can be configured during runtime, was shown in [8], implementing 3x3 convolution layers using the Winograd algorithm. DLAU [7] is another early NN accelerator architecture for FPGAs with the matrix multiplication, addition, and activation fully implemented in logic with three pipelined processing blocks for tile-wise processing. An on-board CPU configures the processing block with parameters and feeds the tiled matrix data. They present a speed-up of 36.1x versus a 2.3 GHz Intel Core2 processor for computing a fully connected network with 32 MAC operations per cycle at 200MHz.

With the upcoming of binary neural networks (BNN) in 2016 [16], allowing MAC computations to be implemented with simple logic, the FINN framework [2] was proposed.

TABLE I
FRAMEWORKS FOR MAPPING NEURAL NETWORKS ON FPGA PLATFORMS

| Framework (year) | Supported networks | Supported precisions | Control (Execution) | Supported HW platforms | Performance (Platform) |
|---|---|---|---|---|---|
| Caffeine (2016) [22] | Same as Caffe | Float, fixed-point | PC via PCIe (layer-wise) | Xilinx FPGA | 1.46 GOPS peak @ 8bit fixed point (Xilinx KU060 FPGA) |
| DLAU (2017) [7] | FC shown | Float | CPU core (layer-wise) | Xilinx Zynq only | 6.4 GOPS peak @ float (Xilinx Zynq 7Z020 SoC) |
| Caffeinated FPGAs (2016) [8] | Only 3x3 conv. in hardware | Float | PC via PCIe (layer-wise) | Xilinx FPGA | 50 GFLOPS on 3x3 convolution only @ 32bit float (Xilinx 7VX690T FPGA) |
| FINN (2018) [2] | FC, CNN | Fixed-point, binary | CPU core (streaming) | Xilinx Zynq only | 397.5GOPS[1] single frame, 2'465 GOPS batch processing, @ binary CNN (Xilinx Zynq ZC706 FPGA SoC) |
| Angel-Eye (2017) [25] | FC, CNN | Fixed-point | CPU core (layer-wise) | Xilinx Zynq only | 137 GOPS, 14.2GOPS/W @ 8bit fixed point (Xilinx Zynq 7Z045) |
| PipeCNN (2017) [24] | FC, CNN | Fixed-point | PC via PCIe (layer-wise) | Intel FPGA SoC | 179 GOPS @ 8bit fixed point (Intel Arria-10 GX1150) |
| fpgaConvNet (2017) [27] | FC, CNN | Fixed-point (16b) | CPU core (streaming) | Xilinx Zynq only | 162 GOPS @ 16bit fixed point (Xilinx Zynq 7Z045) |
| Dedicated YOLOv2 acc. (2020) [32] | YOLOv2 | Fixed-point (8b) | CPU core (streaming) | Intel FPGA SoC | 566 GOPS @ 8bit fixed point (Intel Arria-10 GX1150) |
| MALOC (2018) [28] | FC, CNN | Fixed-point (16b) | CPU core (streaming) | Xilinx Zynq, Virtex-7 | 80.35 GOPS[2] @ 16bit fixed point (Xilinx Zynq 7Z020 FPGA SoC) |
| Eval. Fast Algo. for CNNs on FPGAs (2020) [29] | FC, CNN, others | Fixed-point (16b) | CPU core (layer-wise) | Xilinx Zynq only | 201.1 GOPS[2] @ 16bit fiexed point (Xilinx Zynq ZC706 FPGA SoC) |
| Vitis DPU (2020) [30] | FC, CNN, others | Fixed-point | CPU core | Xilinx Zynq only | 230 GOPS peak @ 8bit fixed point (Xilinx Zynq 7Z020 FPGA SoC) |
| NEURAghe (2018) [31] | FC, CNN | Fixed-point (16b) | CPU core (layer-wise) | Xilinx Zynq only | 169 GOPS @ 16bit fixed point (Xilinx Zynq 7Z045) |
| Unrolling TNN (2018) [33] | FC, CNN | Ternary weight, 16b activations | PC via PCIe (streaming) | Amazon AWS F1 | 2500 GOPS @ ternary weight (Amazon AWS F1) |
| **NN2CAM (this work)** | **FC, CNN** | **Fixed-point & binary** | **Pure logic (streaming)** | **Xilinx FPGA** | **337.8 GOPS average @ binary (Xilinx XC7K325)** |

[1]Computed from their CNN size and the latency, [2]Performance for batch processing only (for a fair comparison, we show their medium-sized FPGA only)

Combined with a streaming architecture, their experiments achieved unprecedented throughputs thanks to the highly parallelizable logic implementation of MAC operations and the small memory space of binary weights allowing to map everything in on-board memory. Additionally, FINN made a first step towards a standalone processing platform as the on-chip CPU core of the utilized Zynq FPGA was used for control purposes. The initial work consisted of three pre-defined networks and a simple compiler that generates HLS code for synthesis using the Xilinx Vivado HLS tool. Recently, significant extensions to the training and quantization-flow of the compiler [23] as well as extensions to other network types were published.

A similar HLS approach was taken in PipeCNN [24], but instead uses Altera's (now Intel) OpenCL flow for high-level synthesis and computing in a layer-wise fashion. Angel-Eye [25] is another framework targeting Zynq-based implementations of CNNs. Their framework consists of a Caffe network-compatible layer-wise quantizer to get fixed-point weights, with a special focus on 8 bit representations, and a compiler that maps the entire network as a standalone engine onto the CPU and the FPGA logic of a Zynq system-on-chip (SoC). The CPU takes care of controlling an FPGA-based CNN accelerator by feeding it with the input data as well as the instructions that are then processed by the accelerator. Another OpenCL-based framework is presented in [26] and their follow-up work [27], reporting a CNN mapping flow for 16 bit fixed point implementations on FPGAs using a latency-driven optimization strategy.

MALOC [28] also features an automated CNN mapping toolchain but implements the entire network on the chip as a pipelined implementation. To enable larger networks, for which all intermediate layers cannot be buffered simultaneously, it employs a tiling mechanism that determines what parts of each intermediate layer shall be buffered. Another framework with an automated mapping toolchain is presented in [29], however, this is focusing on exploiting the efficient implementation of convolution computations using FFT and Winograd transformations. We note that these two implementations use batch processing to keep the utilization in the computation blocks high, which is expected to decrease significantly for single-frame operation where inter-layer dependencies can stall subsequent layer processing as it was shown in the performance of [2].

The latest update of Xilinx' Vitis AI tool provides a deep learning processing unit (DPU) [30], which is an accelerator engine for the Xilinx Zynq platforms, allowing to choose from 8 pre-configured (pre-placed-and-routed) architectures with varying degrees of parallelism. It contains a compiler that maps the network to an efficient instruction set and performs data reuse optimization as well as instruction scheduling. Additionally, they provide an optimizer tool that enables model compression. Vitis AI is part of Xilinx' unified development environment that combines high-level synthesis with multi-processor support and a set of APIs.

NEURAghe [31] is a hardware/software framework for Zynq platforms, exploiting a tight interaction between network-optimized software, running on the hard CPU, and a 16 bit fixed point accelerator implemented on the programmable logic. Another recent work [32] implements a dedicated CNN accelerator for YOLOv2 object detectors using Intel's OpenCL FPGA framework. This accelerator achieves unprecedented throughputs, but only supports 8 bit implementations of YOLO and no other networks.

Tridgell et al. [33] follow a network unrolling approach for ternary networks and additionally exploit sparsity in convolutional layers (using the knowledge of zero weights in the kernels), achieving up to 2.5 TOPS. Due to the unrolling of the network, a high degree of parallelization is achieved but also extreme numbers of FPGA resources are required, which is why it was implemented on a cloud AWS platform and only tested on small input images. A similar unrolling strategy was used in LogicNets [34], encoding the entire network using look-up tables (LUT). Using a new network co-design strategy they can keep the resource utilization low and thus achieve throughputs of more than 8 TOPS on very small low-precision networks. This is unfortunately limited to specialized applications with very small network size like they are used in data communication analytics.

Based on the performance and the flexibility of supported network topologies, [2], [29] and [30] can be considered the state-of-the-art quantized neural network mapping frameworks for FPGA-based hardware acceleration. While some of the other frameworks report up to 3x higher (best in class) throughput results, they are limited in their flexibility in terms of supported quantization levels and supported networks.

Despite the variety of existing frameworks, none of them support the target FastEye camera edge processing use case, requiring flexible and automated NN mapping onto pure-logic FPGA platforms for standalone end-to-end processing, supporting arbitrary-sized CNNs to enable a wide range of applications. While [22] and [8] are designed for Caffe-acceleration on PCs and [8] only supports 3x3 convolutions, all other Xilinx-FPGA-compatible frameworks are limited to Zynq platforms with CPU cores, do not feature standalone processing and only cover limited ranges of precision options. Our work supports the required features for the FastEye camera while achieving 337.8 GOPS, which is comparable to the highest reported performance among the state-of-the-art frameworks.

## IV. NN Mapping Framework

Mapping a trained neural network onto an FPGA platform for on-board inference requires multiple layers of abstraction to be crossed. Machine learning training tools, like Caffe [35], encode the trained network in a set of two file types; one for the (textual) network architecture description and a second one containing the learned parameters in binary format. These files need to be translated into a representation that can be synthesized for the target FPGA platform.

In the following, we describe our NN2CAM framework that automatically maps a trained Caffe-formatted image analysis CNN on an FPGA-based camera as a streaming processing engine as shown in Fig. 5. Starting from the files containing the trained network, the architecture gets extracted using Python scripts, determining all dimensions and layer settings as well as extracting and annotating the trained parameters. After mapping the layer operations to HW-implementable functions (e.g. convolution operation) and allocating the required HW resources to efficiently compute the network, the resulting (high-level) representation gets compiled into a hardware description. From there, the usual FPGA synthesis tools are used to synthesize the system to a bitstream.

Our target system is a high-speed camera, in which the CNN-based image analysis block is instantiated by the proposed framework, automatically mapping arbitrary CNNs to it. However, the mapping flow is generic and can be used for other applications where similar DNN-based image processing on the edge is required.

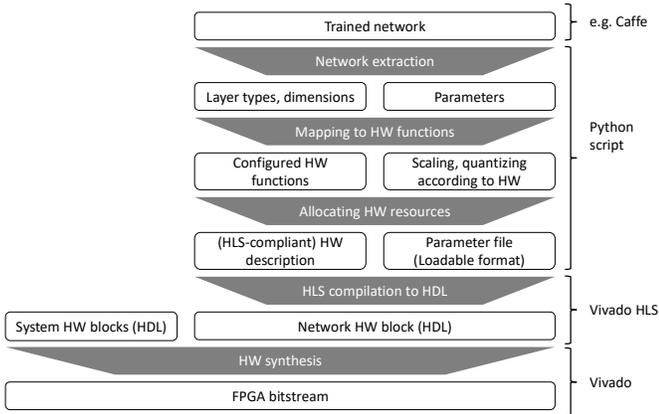

Fig. 5 Framework tool-flow from trained network description to bitstream.

### A. Network extraction

Caffe represents its trained models with 1) a textual prototxt-format file containing the architecture skeleton, describing each layer with its type, dimensions, and connectivity as well as 2) a binary caffemodel-format file that stores all learned parameters like weights and biases. To convert the network into a framework-interpretable format, the architecture description first gets parsed, extracting all layer dimensions, which then allows extracting the parameters and buffering them in a mathematical matrix representation. In this form, the network is essentially a series of 2-dimensional convolution operations on 3-dimensional matrices with known convolution kernels.

### B. Mapping to hardware functions

In the next step, the individual layers are mapped to the available prototype hardware functions. Our tool currently supports the following layer types: 1) FC layers, 2) CNN layers and 3) average-pooling layers which are implemented as high-level C++ functions that are compatible with Xilinx Vivado HLS. All network components like layer types, processing elements, and data handling mechanisms are implemented in this generic C++ format within a set of library files. The compilation script utilizes these functions and parametrizes them during the instantiation. Most of these library functions are based on the FINN framework by Blott et al [1], which was used for implementing BNNs on Xilinx

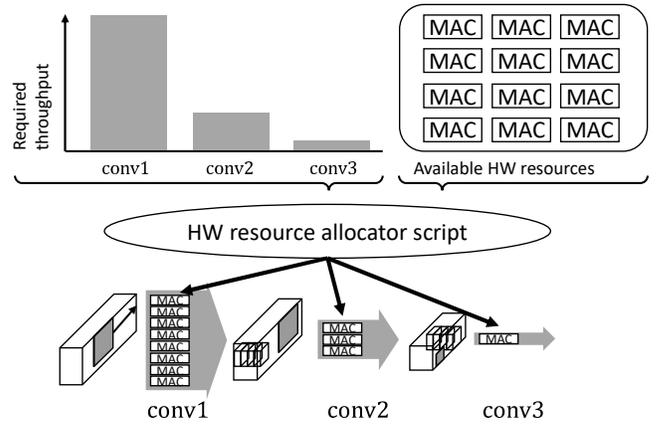

Fig. 6 Computational hardware resource allocation to ensure a continuous data processing stream across all layers.

Zynq FPGAs. As FINN was initially focusing on BNNs, it did not directly support arbitrary fixed-point precision networks nor implement input activation padding or ReLU activation functions, such that some of the libraries had to be extended. We list the main extensions to the original FINN library:

- Extended with arbitrary fixed-point precision.
- Extended sliding window function with side padding and fixed-point precision support.
- Added a function for ReLU activation support.
- Extended MAC implementation with fixed-point precision support.
- Mapping tool: The FINN framework did not feature a mapping tool but only supported a set of pre-configured networks.

Looking at the processing of a CNN layer, input activations from a sliding window position need to be multiplied and accumulated with a learned weight kernel. The MAC operations of each window position perform a dot product between the two matrices, which can be broken down into a sequence of MAC operations. The same operations are repeated for each window position on the input feature map. The data path can thus be split into two stages: 1) the sliding window generation that buffers the input data to provide the correct input activations (window) and 2) the MAC processing that performs the actual computation of the data. Based on the extracted layer dimensions, the prototype layer functions can be configured for each layer.

As explained earlier, FC layers can be interpreted as a subset of CNN layers where the kernel has the same size as the input layer, making the sliding window operation needless. Average pooling layers are also similar to CNN layers but work in a channel-wise manner with a homogeneous weight kernel that averages over the input activations of each input channel separately. To avoid multiplications, the input activations can be summed up and the sum divided by the number of kernel elements per channel. This framework assumes that the parameters of the trained network have already been quantized during training and thus directly quantizes extracted parameters to the fixed-point precision that is specified in the model.

## C. Allocating and balancing hardware resources

After mapping the network to hardware-representable functions, the internal network representation is functionally complete. However, due to the arbitrary size of each layer, the layer processing time can vary significantly across the layers, requiring a processing throughput balancing between them. This is necessary because each layer in the streaming architecture is waiting for data from its previous layer, stalling the following ones if insufficient input activations are supplied. Unbalanced layers therefore negatively impact the temporal processing element utilization in following layers, leading to a low overall throughput. Fig. 6 visualizes this problem, comparing each layer's processing workload in terms of MAC operations per output to achieve a continuous data flow in all layers. The assumed 3-layer network shows higher computational loads in the first layers, decreasing towards the output.

Because these relations are highly network-dependent, it is necessary to balance the computational resources for each network individually, taking both the network and the available resources into account. By distributing the available hardware resources of the FPGA to the layers according to the distribution of required throughput across all layers, a network-wide balance of layer throughputs can be found. A balance on a specific layer is achieved when the time for computing all output channels of a certain x/y position equals the time for computing and buffering sufficient input activations to move the sliding window to the next position in the x/y-plane as depicted in Fig. 7.

We compute this balance by quantifying the average rate of input activations (see equation (3)) required to compute the average rate of output activations (see equation (4)) and compute the ratio $R_i$ between them, as shown in equation (5). These equations depend on the number of processing elements as well as the time for computing one kernel, but all implementation-specific parameters cancel out in equation (5). Knowing $R_i$ for each layer $i$, allows to determine the balanced distribution of MAC units across the layers, ensuring that the input activation rate of each layer matches the output activation rate of the previous one.

Equation (6) formulates this distribution $D_{MAC}$, by computing the normalized product of all $R_i$ for each of the separate layers. From equation (4) and (6) follows, that the product of $\#PE \cdot \#SIMD$ MAC units can be found for each layer $i$ by multiplying the available number of processing elements (e.g. DSPs) with $D_{MAC}(i)$. Binary MAC units do not consume any DSP elements, making the search for the available number of processing elements an iterative process that depends on available logic elements.

Our framework takes the number of available DSP resources as well as the number of block RAM (BRAM) into account. Resources already used by the hosting application (the image acquisition data path in our camera application), are deducted prior to the distribution. The streaming architecture supports the following two kinds of processing parallelism within a layer:

1. Output channel parallelism: multiple output channels are being processed at the same time. Each output channel is processed in a separate processing element (PE). Input activations can thus be reused across all PE's and each PE independently accumulates the results of its multiplications. The number of PE's is limited by the number of output channels in the layer. Additionally, each PE is independently accessing parameters from the memory, requiring a dedicated BRAM-macro instance per PE.
2. Input parallelism: multiple input activations in each PE are being processed in parallel. We refer to this as single instruction, multiple data (SIMD) parallelism, as it was called in the FINN framework [2]. To simplify the implementation, the number of parallel input computations (#SIMD) is restricted to integer divisors of the number of elements in the kernel.

$$in_{rate} = \frac{\#in\ (s_y\ rows)}{t(1\ output\ row)} = \frac{s_y \cdot X_{in} \cdot C_{in}}{\frac{X_{in}}{s_x}\left(\frac{C_{out}}{\#PE} \cdot \frac{t(1\ kernel)}{\#SIMD}\right)} \quad (3)$$

$$out_{rate} = \frac{\#out}{t(1\ output\ pos.)} = \frac{\#PE}{\left(\frac{t(1\ kernel)}{\#SIMD}\right)} = \frac{\#PE \cdot \#SIMD}{t(1\ kernel)} \quad (4)$$

$$R_i = \frac{out_{rate}}{in_{rate}} = \frac{C_{out}}{s_y \cdot s_x \cdot C_{in}} \quad (5)$$

$$D_{MAC}(i) = \frac{\prod_{k \leq i} R_k}{\sum_i (\prod_{k \leq i} R_k)} \quad (6)$$

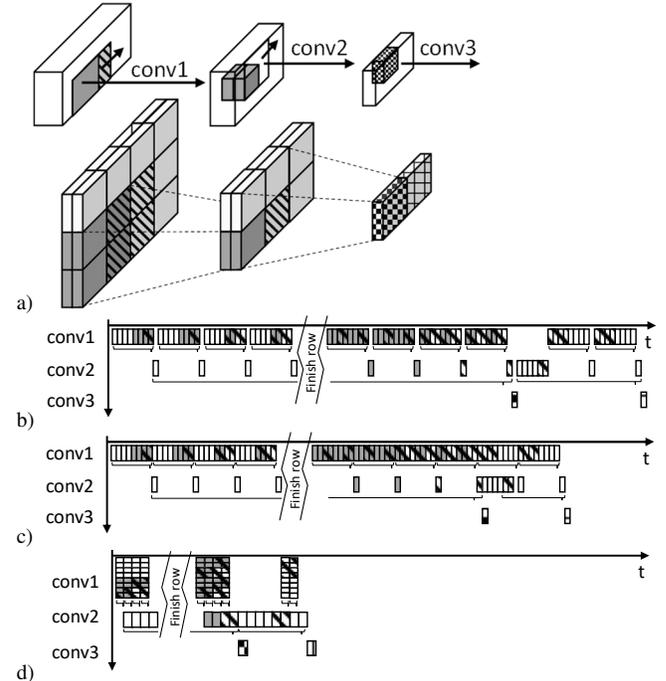

Fig. 7 a) Visualization of connected regions across convolutional layers of an example CNN. All convolutions have a stride of 1, indicated with the two neighboring window positions (gray, striped). b) Sequence of MAC operations on each layer. MAC computations are marked according to the input region they belong to. Note that convolutions 2 and 3 are stalled for a substantial part of the time, waiting for dependencies to become available. c) Same MAC sequence with layer pipelining implemented. d) In addition to the pipelining, this sequence includes 8x parallelization in convolution 1, leading to a non-stalling (balanced) stream of MAC operations in the second layer.

## D. HLS compilation to hardware representation

The last step of the mapping process consists of translating the framework-internal high-level representation into a hardware description. This framework first compiles the mapped network into an intermediate C++-based high-level representation that is compatible with Xilinx Vivado HLS. Thus, it can be automatically synthesized into HDL using Vivado HLS, avoiding user-interaction with the more complex HDL code. The synthesized block features two standard advanced extensible interfaces (AXI4), which allows to automatically map the block into an adapter block that contains all utility blocks to load the input image and all the parameters during startup through the AXI interfaces. This includes an image buffering and loading block, a result extraction and buffering block as well as a control block as shown in Fig. 8. These utilities simplify the instantiation of the accelerator in any FPGA design, only requiring two interfaces: 1) an image line input and 2) a result (output layer) interface.

## V. STREAMING PROCESSING ARCHITECTURE

The layered structure of a CNN implemented as a streaming process imposes sequential processing of activations as they are crossing intermediate layers. Each x/y-position of a certain layer depends on activations from a limited (kernel-sized) input region on its preceding layer as shown in Fig. 7. Thus, the output activation computation can only be completed once all input activations are available, implying that the following layer's computations will be stalled if the input activations are not computed fast enough. Fig. 7 a) and b) visualize this problem, where the computation of the first convolutional layer conv1 is stalling the processing of the second convolution. To process MAC operations in parallel across all layers, pipelining of the layer computations is used, allowing each layer to compute new activations while the following layer is still busy processing the previous ones.

Due to these challenges, the framework is mapping the network to a systolic streaming architecture, instantiating separated processing blocks for every network layer, allowing layer-pipelining and layer-individual MAC parallelization. The system features two types of computational parallelism:

1. Inter-layer parallelism: layer-pipelining allows each layer to optimize its compute utilization by virtually removing the dependency between layers through time-shifted processing, as shown in Fig. 7. c). This pipelining parallelism makes it possible to process layers in parallel even though the result of each layer might be used in the following one.
2. Intra-layer parallelism: Each layer processing block can parallelize its computations independently. This allows the computational power to be optimized for the layer dimensions and the input requirements of the following layer, as shown for convolution 1 in Fig. 7. d). We implement two mechanisms for intra-layer parallelism, namely output channel parallelism and intra-kernel parallelism.

Fig. 8 shows the block diagram of the targeted camera system including the accelerator block in which the neural network is instantiated by the framework. The input image from the (camera) application is stored in a buffer that is accessible by the accelerator. The two standard AXI4 ports communicate with the accelerator, the first one accesses the data while the second one is used for controlling and parameter loading.

Buffered image data is automatically read by the accelerator through the data interface and converted internally into a sequence of streams, passing through the network layers. Each such stream consists of a FIFO that buffers data until they are requested by the subsequent layer for processing. Figure 9 shows the implementation of a three-layer network using layer processing blocks which compute the instantiated layer type using locally stored weights and biases. Each such layer operates in parallel to all others (inter-layer parallelism). The parameters are loaded through the control interface during the initialization of the accelerator. Results from the last layer are streamed out through the data interface and buffered in an accelerator-external result buffer.

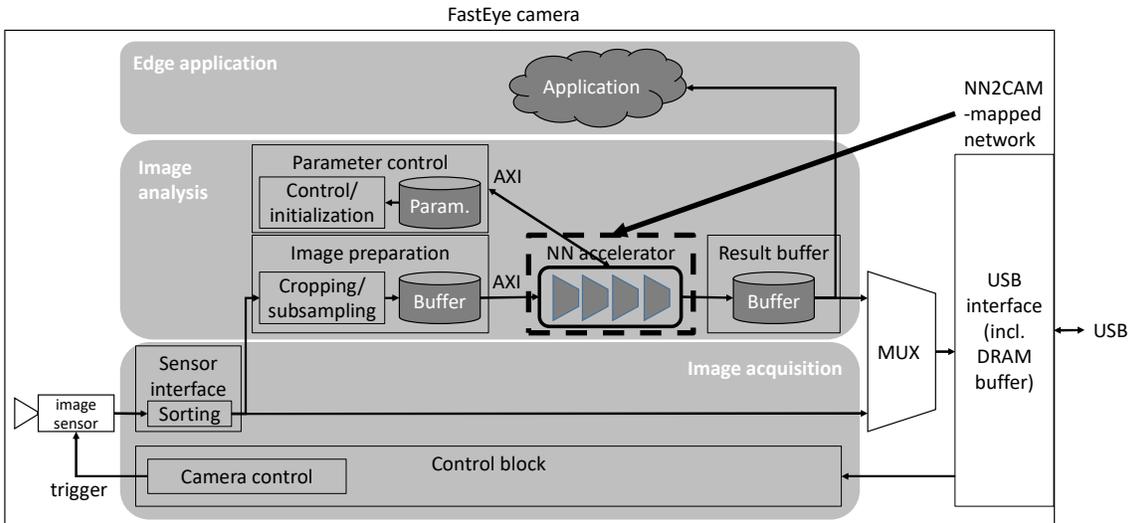

Fig. 8 Block diagram of the implemented FastEye camera system with automatically mapped NN accelerator block.

Once the network inference is complete, the application can access the results from the buffer and trigger the next execution. A more detailed look into the block reveals the NN-specific mathematical operations, namely the MAC operation that multiplies and accumulates inputs with weights, and a sliding window input generator that generates the correct input activation access pattern for 2D convolution. The control block contains all necessary utilities and state machines for controlling parameter loading and system state control. Intra-layer parallelism is provided through instantiation of multiple PEs, allowing to process multiple output channels in parallel, reusing the same input data. Each of these PEs can be configured to additionally process multiple inputs of its kernel-computation in a SIMD fashion. Instead of processing a single input activation from the input stream per cycle, multiple ones are fetched and processed in parallel, adding another dimension of parallelism.

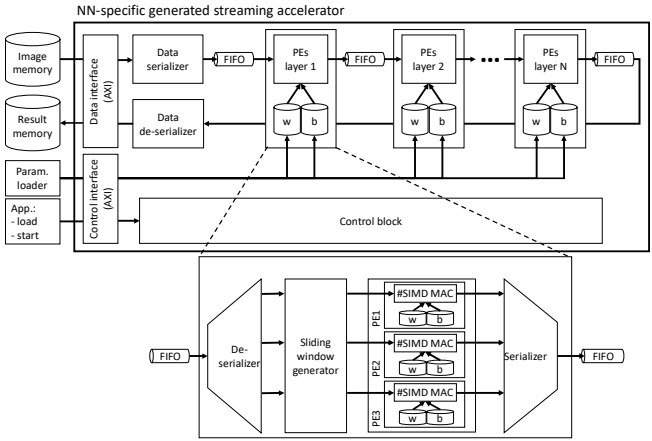

Fig. 9 Block diagram of an implemented streaming architecture with data streams connecting the layer processing blocks. Below, a more detailed view of a layer processing block is shown, consisting of a stream interpreter that feeds an optional sliding window generator, followed by MAC processing units (with local weight and bias memories) and a data serializer that generates the output stream.

### A. Sliding window generator

The 2-dimensional convolution in CNNs requires the convolution kernel-sized input window to be slid over the input feature map to forward the input features of each window position along with the corresponding kernel weights to the processing block for performing the MAC operations. Because the input window is only moved by a few input elements in x or y direction (defined by the stride parameters $s_x$ and $s_y$), most of the window data can be reused from the previous window, avoiding the same input data to be read multiple times from the preceding layer. This is achieved by instantiating a buffer that holds the number of input rows needed for computing 2 output lines. While the input data for the first output line is being used for output computations, the remaining buffer is continuously being filled at the same time, such that the next output row can be started directly after the current row is completed. Figure 10 visualizes this concept.

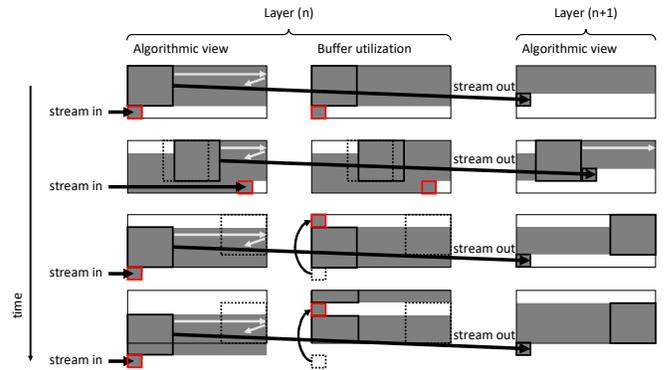

Fig. 10 Temporal evolution of activations data buffer states of two subsequent layers showing data reuse and continuous filling for the next row.

### B. Processing element

The PEs are implemented as simple arithmetic units because the same data order is ensured in both the input activations stream and in the kernel weights allocation within the memory. This avoids having to rearrange data and limits their task to multiplying and accumulating the incoming data in the order they arrive (input activations from the stream, weights from its memory), adding the bias at the end of each kernel (after finishing all KERNEL_SIZE inputs) and then applying the activation function. By running this state machine for every element in the output feature map completes the computation of a layer. The responsibility for feeding the PEs with the correct data is thus shifted to the sliding window generator.

The data flow of a layer is depicted in Fig. 11, showing the data handling from the input activation stream, passing through the sliding window data generation, the processing elements, and finally ending up in the output stream that continues to the next layer. Data in streams are ordered channel-first, streaming all input channels of an x/y position, followed by all channels of the neighboring x/y position in the x-direction and after completing a row continuing with the next row in the y-direction. All PEs receive the same input activation data but compute them for a different output channel (output channel parallelism). In every cycle, each PE computes #SIMD input channels, such that computing a single output element requires KERNEL_SIZE/#SIMD cycles (intra-kernel parallelism). If there are more output channels than allocated PEs, the following #PE output channels get computed afterward (before moving on in x/y-direction). This creates the correct output data ordering as explained before, allowing the next layer to be processed in the same manner. Each PE's parameter memory is mapped to match this activations sequence with every logical memory address containing #SIMD weights, making the weights of a single kernel to appear in a sequence, followed by the next kernel to be processed by the PE ($C_{out}$ modulo #PE).

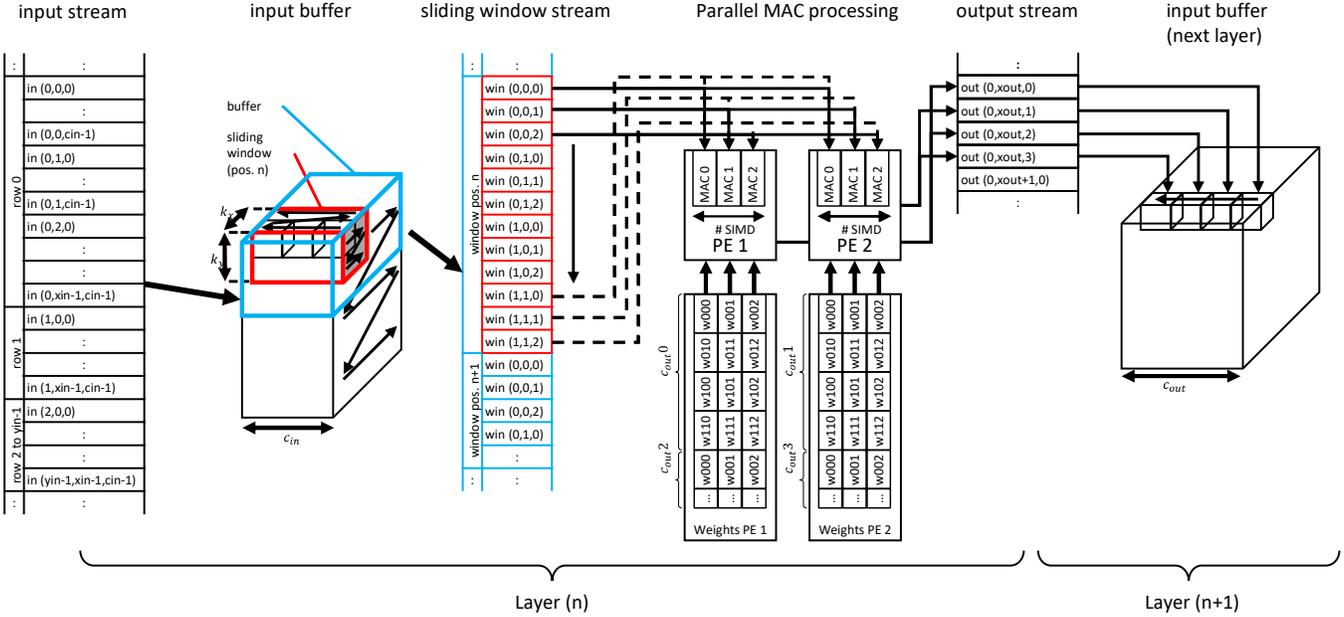

Fig. 11 Data-stream order and parallelism in the streaming pipeline while computing a 2x2 convolution on a 3-channel input/4-channel output layer. Input activations are fed in as a stream, while weights are read from each PE's dedicated memory. Bias/threshold values and the output activation function are omitted for simplicity.

*C. Precision*

The computational precision used for processing neural networks has been investigated intensively over the past few years as aggressive quantization was shown to reduce the required hardware resources and the power consumption. As a result, a wide range of activation and parameter precision options, ranging from fully binary precision in BNNs to 16 bit values or higher, are used across different network implementations. BNNs simplify the dominating multiply-accumulate (MAC) operations to simple XNOR logic with appended bit-counting (popcount) [2]. This is highly advantageous for FPGAs as the memory needs are reduced and XNOR multipliers with popcount adders can be efficiently implemented in pure logic, minimizing the use of limited hardware resources like DSP blocks.

Because of these advantages for FPGA implementations, a unique feature of our framework is to allow arbitrary precision of weights and activations. The precision setting can be applied to each layer separately, allowing, for instance, networks with low precision feature extraction in their first layers to be implemented along with higher precision classifiers in their last layers. Support for higher precision computations is often necessary in the first and last layers to achieve high accuracy in highly quantized networks [36]. Apart from the fixed-point precision computations, that get synthesized into DSP-based computation blocks, the framework features the mentioned XNOR-based implementation for fully binary computations. The framework automatically determines these layers from the network description and selects the XNOR implementation from the MAC library. Whenever a value is converted into a fixed-point representation of different precision, e.g. from a higher precision MAC result or from a binary (low precision) input into a fixed-point activation, the fractional data are directly quantized through truncation and a saturation mechanism ensures that the minimum/maximum representable value is used in case the target data type is not sufficiently wide. This fixed-point data type is provided by Vivado HLS (type ap_fixed).

*D. Portability*

To allow the presented neural network mapping framework to be used on a wide variety of FPGA models, including the Kintex-7 FPGA on our target platform, we avoided the use of family-specific resources like hard CPU cores. Most other frameworks require such cores for control purposes, limiting the FGPA selection to Zynq-based platforms (for Xilinx-compatible tools).

Due to the use of vendor-specific HLS libraries and thus also IP blocks, an HLS-compiled IP block cannot directly be transferred to other HLS tools of different vendors. However, within products of the same manufacturer, most HW platforms can be targeted through compiler options in the HLS flow. Other HLS tools have similar design flows and could possibly be used in the same fashion as presented here (e.g. Intel high-level synthesis compiler [37]).

## VI. EXPERIMENTS AND RESULTS

We evaluate the performance of our approach by employing the proposed framework for implementing a set of neural networks on our target platform, an FPGA-based FastEye high-speed camera [9]. In Table II we report the FPGA resource utilization as well as the algorithmic performance and power consumption.

The FastEye platform features a 1 megapixel high-speed image sensor that is directly connected to a Xilinx XC7K325 FPGA. Its wide parallel interface to the sensor enables fast data extraction and pixel sorting to prepare image data for transmission through a USB interface. Fig. 8 shows these interfaces to the sensor and the USB controller at the input and the output of the system, respectively. A control block takes care of configuring the image sensor and forwards commands sent via USB to the other blocks. The original camera implementation directly sends acquired and sorted image data to the USB interface from where the images can be accessed by a host computer.

We implement the following use-cases with the presented NN2CAM framework on the FastEye camera and report the FPGA resource utilization as well as the measured power consumption. The accelerator is clocked at 100MHz and reads the acquired image, cropped to 28x28 - 640x640 pixels, from a block RAM buffer memory (128 BRAM blocks).

*A. 640x640 pixel optical character detection and recognition*

Many industrial quality control applications require number and text recognition to be performed to identify objects using various kinds of labels. The position of these labels is usually variable, making it necessary to analyze a larger field of view (e.g. 640x640 pixels) at the frame rate required to keep up with the speed of conveyor belts and other movements induced during manufacturing. More detailed analysis (e.g. high accuracy network analysis with higher precision arithmetic) can then be performed on smaller regions of interest. Alternative systems with cloud- or GPU-accelerated processing require costly high-end infrastructure like high-speed video communications and network connections.

Optical character recognition (OCR) based on the MNIST dataset, containing small 28x28 pixel images of hand-written digits, is often used as an example to show a basic functionality of a neural network. However, this dataset is not very useful for real-world applications, such that we augmented the dataset to account for illumination change, resilience to background noise. Using this MNIST-based OCR (M-OCR) dataset, we train two 5-layer CNNs for classifying digits as shown in Fig. 12. The network output dimension is 1x1x11-78x78x11, for input dimensions 28x28-640x640, respectively, with each output channel representing one of the 10 possible numbers or the undefined class. The CNN is quantized to either 16 bits or to fully binary representations, while the output layer is always 16 bit wide. Using the binary implementation, the FPGA resource utilization is substantially reduced compared to the 16 bit implementation of the same network, allowing to implement a higher degree of parallelism, achieving 17x-76x higher throughputs. It must be noted that the large buffers for the input image and the results already amount to 128 BRAM blocks, for the largest resolution. This limits the maximum implementable parallelism in the 16 bit implementation (higher parallelism requires more BRAM blocks as each PE requires a separate BRAM), limiting the throughput and thus keeping the latency high. Fig. 13 illustrates this by comparing the throughput with the image size. While the throughput of binary implementations increases with the image size (the latency of small images is dominated by the filling of computing pipelines), the throughput for 16 bit implementations rapidly drops to a low level for image sizes above 28x28 as parallelism becomes limited by the lack of available BRAM resources (used up by the increasing FIFO buffers). Reducing input and result buffers by directly feeding the image data to the accelerator without buffering it and directly transmitting the results through USB could reduce this limitation in the future.

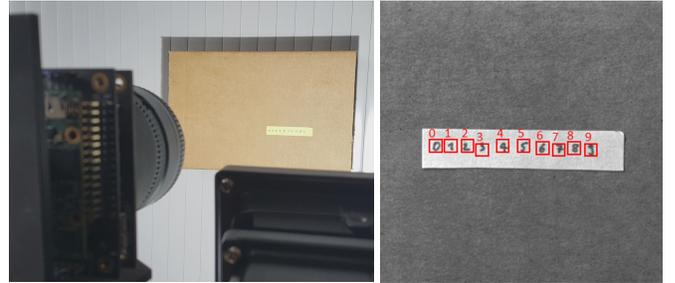

Fig. 12 640x640 pixel M-OCR edge processing on FastEye high-speed camera.

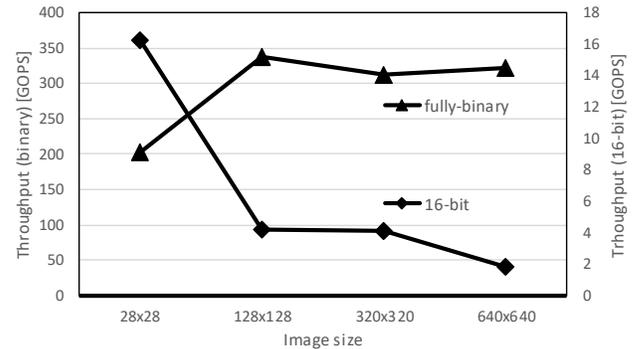

Fig. 13 Throughput for different image sizes and arithmetic precisions.

TABLE II
EXPERIMENTAL RESULTS

| Implementation | Network | | | | Comp. performance | | FPGA utilization | | | | Sys. power [W] |
|---|---|---|---|---|---|---|---|---|---|---|---|
| | Architect. | Size (MOP) | Precision | Accuracy | Latency [ms] | Throughput [GOP/s] | LUT | FF | DSP | BRAM | |
| Camera only | - | - | - | - | - | - | 28.6k (14%) | 82.4k (20%) | 8 (1%) | 12 (3%) | 12.6 |
| 28x28 OCR | 5-layer CNN | 1.5 | Fully binary | 92.5% M-OCR | 0.007 | 203.0 | 167.9k (82%) | 196.5k (48%) | 36 (4%) | 328.k (74%) | 13.2 |
| 128x128 OCR | 5-layer CNN | 147.0 | Fully binary | 92.5% M-OCR | 0.458 | 321.0 | 187.9k (92%) | 203.4k (50%) | 71 (8%) | 331.5 (74%) | 13.1 |
| 320x320 OCR | 5-layer CNN | 1054.3 | Fully binary | 92.5% M-OCR | 3.375 | 312.4 | 156.9k (77%) | 177.1k (44%) | 79 (9%) | 361.5 (81%) | 13.2 |
| 640x640 OCR | 5-layer CNN | 4406.6 | Fully binary | 92.5% M-OCR | 13.311 | 337.8 | 188.1k (92%) | 203.6k (50%) | 80 (10%) | 432.5 (97%) | 13.1 |
| 28x28 OCR | 5-layer CNN | 1.5 | 16-bit | 98.9% M-OCR | 0.092 | 16.3 | 99.6k (49%) | 152,6k (37%) | 101 (12%) | 435 (98%) | 13.5 |
| 128x128 OCR | 5-layer CNN | 147.0 | 16-bit | 98.9% M-OCR | 35.001 | 4.2 | 100.1k (49%) | 152.9k (38%) | 136 (16%) | 444 (100%) | 13.5 |
| 320x320 OCR | 5-layer CNN | 1054.3 | 16-bit | 98.9% M-OCR | 257.146 | 4.1 | 96.4k (47%) | 148.6k (36%) | 100 (12%) | 443 (100%) | 13.5 |
| 640x640 OCR | 5-layer CNN | 4406.6 | 16-bit | 98.9% M-OCR | 2498.110 | 1.8 | 96.6k (47%) | 148.5 (36%) | 93 (11%) | 436.5 (98%) | 13.5 |
| 640x640 face det. | 9-layer CNN | 420 | Mixed 1/16-bit | 92.3% FDDB | 28.027 | 15.0 | 112.2k (55%) | 203.0k (50%) | 470 (56%) | 417 (94%) | 13.2 |

*B. 640x640 pixel face detection*

To demonstrate that also more complex tasks, like multi-scale face detection on 640x640 pixel images, can be implemented using this system, we trained and mapped a 9-layer CNN, achieving 92.3% accuracy on the FDDB benchmark, onto the camera. The network is quantized to 16 bit activations and has binary weights in its first 7 layers (last 2 layers have 16 bit weights). Even though its input size of 640x640 pixels is equal to the large OCR network, the combination of binary weight precision and a coarser detection granularity reduce the required BRAM resources and enable a frame rate of more than 40 FPS, achieving a throughput comparable to the small resolution OCR use case at 16 bit precision.

Comparing the performance of NN2CAM with the prior work in Table I shows that even though it implements the complete end-to-end system, NN2CAM achieves comparable or higher throughput performance in its binary use cases. The binary implementation of use-case A achieves on-par throughput compared to FINN's fully-binary CNN accelerator [2] for batch size 1. Their experimental setup using stored test images additionally allows frames to be processed in batch mode, which enables 6x higher throughput due to better resource utilization. In a real setup, where camera images must be processed frame-by-frame, this operating mode is not possible. The same reduction in throughput for single frame operation is expected for [28] and [29], while they additionally only support 16 bit fixed point arithmetic. However, these two frameworks achieve a higher throughput for 16 bit implementations. Compared to the 8 bit precision implementations [24], [25] and [30], NN2CAM achieves 1.5x-2.5x higher throughput for binary implementations but up to 14x lower throughput for 16 bit implementation. However, [24] only supports a set of predefined networks, making it less flexible. While [8] achieves 3x higher throughput than our 16 bit implementations, it is limited to 3x3 convolutions and does not support standalone operation. On the lower end of SoA performance comparison, [7] and [22] report peak performances in the range of our lowest average performance but more than 2 orders of magnitude lower than the full-binary used cases presented here.

## VII. EXTENSIONS AND LIMITATIONS

The presented framework supports CNN and FC networks with arbitrary precisions and the standard ReLU activation. Due to the HLS-based implementation, the library can be extended using high-level C++ functions, which allow adding new activation functions or layer types without having to write HDL code. The framework is mainly constrained by the available resources of the target FPGA platform (e.g. available memory and number of DSP units), limiting the maximum network size and throughput.

We envision the support for additional resource-efficient low-precision arithmetic operators (e.g. for ternary weight MAC), that can be implemented using similar resource-saving logic as for the presented fully binary layers. Additionally, future versions could implement residual layers and separable convolutions, allowing MobileNets to be automatically mapped onto FPGA using this framework.

## VIII. CONCLUSION

We presented NN2CAM, an end-to-end framework for automatically mapping trained quantized neural networks onto FPGA-based edge processing devices and show experimental results on a FPGA-based high-speed camera system. The experiments showcased its support for arbitrary fixed-point precisions with an efficient XNOR-implementation for fully binary layers as well as a computational resource balancing mechanism for effective parallelization within each layer and across the entire network. Utilizing the proposed framework for implementing a customized network architecture on the camera simplified the development process to compiling the trained network into an IP block and instantiating it in the camera firmware. In contrast to other implementations, this framework does not require a powerful CPU core on the FPGA and can thus be implemented on a wide range of FPGA platforms.

Our edge processing experiments implemented on the camera show computational throughputs of up to 337 GOPS, which is SoA performance for single-frame inference, and provide the flexibility of running networks at various arithmetic precisions on the example of implementations with fully-binary, 16 bit and mixed binary/16 bit precisions.